\documentclass[epj,a4paper]{svjour}
\usepackage{graphicx}
\usepackage[utf8]{inputenc}  
\usepackage[T1]{fontenc}
\usepackage{color}
\usepackage{amsmath}
\usepackage{dcolumn}
\usepackage[sort&compress]{natbib}
\setcitestyle{numbers,,open={[},close={]}}
\usepackage{setspace}
\usepackage{tabularx, dcolumn}
\usepackage{booktabs,array}
\usepackage[colorlinks,linkcolor=blue,citecolor=blue,urlcolor=blue,bookmarks=false,hypertexnames=true]{hyperref}
\usepackage[flushleft]{threeparttable}
\usepackage{verbatim,multirow}
\usepackage{siunitx}
\usepackage{upgreek}

\newcolumntype{d}[1]{D{.}{.}{#1}}
\begin{document}

\title{The Felsenkeller shallow-underground laboratory for nuclear astrophysics}
\author{Daniel Bemmerer\inst{1} \thanks{email: d.bemmerer@hzdr.de},
Axel Boeltzig \inst{1}, 
Marcel Grieger \inst{1}, 
Katharina Gudat \inst{1}, 
Thomas Hensel \inst{2,1}, 
Eliana Masha \inst{1}, 
Max Osswald \inst{1,2}, 
Bruno Poser \inst{1,2}, 
Simon Rümmler \inst{1,2},
Konrad Schmidt \inst{1},
José Luis Taín \inst{3},
Ariel Tarifeño-Saldivia \inst{3,4},
Steffen Turkat \inst{2},
Anup Yadav \inst{1,2},
Kai Zuber \inst{2}
}
\institute{
Helmholtz-Zentrum Dresden-Rossendorf, Bautzner Landstraße 400, 01328 Dresden, Germany \and
Technische Universität Dresden, Institut für Kern- und Teilchenphysik, Zellescher Weg 19, 01069 Dresden, Germany \and
Instituto de Física Corpuscular, CSIC and Universitat de Valencia, E-46980 Paterna, Spain \and
Institut de Tècniques Energètiques (INTE), Universitat Politècnica de Catalunya (UPC), Av. Diagonal 647, Barcelona, Spain
}
\date{\today}

\abstract{
In the Felsenkeller shallow-underground site, protected from cosmic muons by a 45\,m thick rock overburden, a research laboratory including a 5 MV Pelletron ion accelerator and a number of radioactivity-measurement setups is located. The laboratory and its installations are described in detail. The background radiation has been studied, finding suppression factors of 40 for cosmic-ray muons, 200 for ambient neutrons, and 100 for the background in germanium $\gamma$-ray detectors. Using an additional active muon veto, typically the background is just twice as high as in very deep underground laboratories. The properties of the accelerator including its external and internal ion sources and beam line are given. For the radioactivity counting setup, detection limits in the 10$^{-4}$ Bq range have been obtained. Practical aspects for the usage of the laboratory by outside scientific users are discussed.
}


\authorrunning{D. Bemmerer {\it et al.}}

\maketitle


\section{Introduction}

Nuclear astrophysics strives to understand the origin of the chemical elements, through experiments, stellar modeling, and astronomical observations \cite{Iliadis15-Book}. One important aspect of this field is the experimental determination of the cross sections of astrophysically relevant nuclear reactions. For charged particle induced cross sections, the absolute values of their cross sections are low, and their study entails low-background sensitive measurements.

The technique of low-background cross section measurements at ion accelerators placed in an underground laboratory has proved especially powerful in this regard. In particular, there has been considerable progress in the knowledge of the nuclear reactions of solar and stellar hydrogen burning \cite{Adelberger98-RMP}, driven to a large extent by underground ion accelerators \cite{Adelberger11-RMP,Acharya24-RMP}.

The underground ion accelerator technique was pioneered by the LUNA collaboration, working in the INFN Laboratori Nazionali del Gran Sasso (LNGS) in Italy, deep underground below 3800\,meters water equivalent (m.w.e.) of rock. A large number of relevant nuclear reactions was studied in this unique low-background environment \cite{Broggini10-ARNPS,Broggini18-PPNP,Aliotta22-ARNPS}. LUNA has used three ion accelerators, namely a 50\,kV machine \cite{Greife94-NIMA}, now decommissioned, a 400\,kV machine for $^1$H and $^4$He beams \cite{Formicola03-NIMA}, running since 2001, and the 3.5\,MV Bellotti Ion Beam Facility (formerly LUNA-MV) accelerator for $^1$H, $^4$He, and $^{12}$C beams \cite{2023FrP....1191113J}, commissioned in 2023.

Based on these successes, other underground ion accelerators have recently been installed in several countries. They include the CASPAR 1\,MV accelerator in the Sanford Underground Research Facility in the Homestake mine, South Dakota \cite{Robertson16-EPJWOC}, the JUNA 0.4\,MV accelerator at Jinping Underground Facility in China 
\cite{Liu16-ScienceChina}
and the Felsenkeller 5\,MV underground ion accelerator in Germany (this work).

While the LUNA / LUNA-MV, CASPAR, and JUNA accelerators are all located in deep underground laboratories with rock overburdens in excess of 1000\,m, at Felsenkeller the rock cover is only 45\,m thick. For the present purposes, rock overburdens in excess of 1000\,m are called "deep", while overburdens more similar to the 45\,m of Felsenkeller are called "shallow". 

The usefulness and suitability of the deep-underground accelerator labs is clear from the experimental results \cite[to cite a few]{Broggini10-ARNPS,Broggini18-PPNP,Aliotta22-ARNPS,Zhang22-Nature}.
The suitability of Felsenkeller for low-background nuclear astrophysics work, however, had to first be proven with detailed background studies. These studies have been carried out in recent years and address several aspects: First, the muon flux and angular distribution \cite{Ludwig19-APP}. Second, the neutron flux and energy spectrum \cite{Grieger20-PRD}. Finally and most importantly, the remaining background in $\gamma$-ray detectors with and without muon veto, both in the range of radiative capture $\gamma$ rays \cite{Szucs19-EPJA} and in the range of natural radionuclide $\gamma$ rays \cite{Turkat23-APP}, with comparisons between overground, shallow-underground, and deep-underground laboratories \cite{Szucs12-EPJA,Szucs15-EPJA,Szucs19-EPJA,Turkat23-APP}.

The aim of the present work is to provide a comprehensive description of the Felsenkeller laboratory and installations, and to present some new data giving an impression of the lab's capabilities. This is timely because of the recent publication of a re-measurement of the astrophysically relevant $^{12}$C($p,\gamma$)$^{13}$N cross section at Felsenkeller \cite{Skowronski23-PRC}, overlapping in energy with a related measurement at LUNA \cite{Skowronski23-PRL}. Further Felsenkeller-based measurements on Big Bang and hydrogen burning reactions are under analysis \cite{Turkat23-PhD}.

This work is structured as follows. The laboratory site and architecture are described in section \ref{sec:Laboratory}. Section \ref{sec:Background} reports on the various background measurements, including also new data updating and complementing Refs. \cite{Ludwig19-APP,Grieger20-PRD,Szucs19-EPJA,Turkat23-APP}. The Pelletron ion accelerator and its internal and external ion sources are described in section \ref{sec:Accelerator}. The in-beam $\gamma$-ray spectroscopy setups are discussed in section \ref{sec:InbeamSetup}. In section \ref{sec:RadioactivitySetups}, the several offline $\gamma$ counting setups at Felsenkeller are shown. A summary is given in section \ref{sec:Summary}.

\section{Description of the laboratory}
\label{sec:Laboratory}

\subsection{Location and overburden}

\begin{figure}[tb]
\includegraphics[width=\columnwidth]{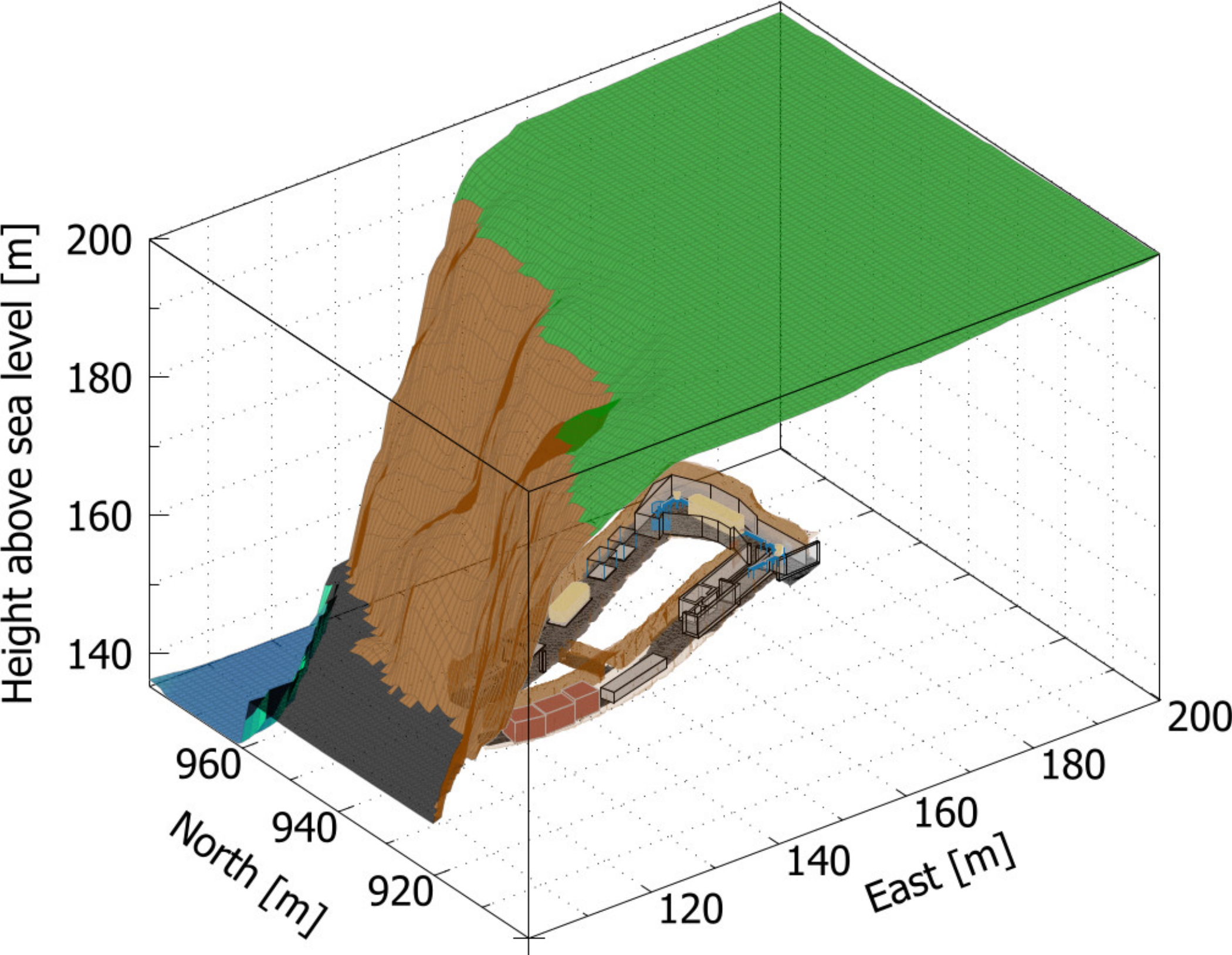}
\caption{\label{fig:BirdsEye} Bird's eye view of the Felsenkeller tunnels with the main installations. The locations are referred to a a reference point at 51$^\circ$00'50" North, 13$^\circ$42'05" East. Elevations are given above sea level.}
\end{figure}

The Felsenkeller shallow-underground laboratory (Figure \ref{fig:BirdsEye}) is located in Dresden, Germany, and is shielded by a rock overburden of 45\,m (140\,m.w.e.) \cite{Ludwig19-APP}. The laboratory  is part of a former industrial site and built into the tunnels VIII and IX of this area. The surrounding hornblende monzonite \cite{Paelchen08-Book} contains the natural radionuclides $^{238}$U and $^{232}$Th with specific activities of 130(30)\,Bq/kg and 170(30)\,Bq/kg respectively \cite{Grieger20-PRD}. The floor of the tunnel laboratory is located at about 140\,m above sea level. 
The laboratory has been developed jointly by the two institutions Helmholtz-Zentrum Dresden-Rossendorf (HZDR)\footnote{Web site https://www.hzdr.de } and Technische Universität Dresden (TU Dresden)\footnote{Web site  https://www.tu-dresden.de}.

\subsection{Buildings and passive shielding inside the tunnels}
\label{subsec:PassiveShielding}

\begin{figure*}[tb]
\includegraphics[width=\textwidth]{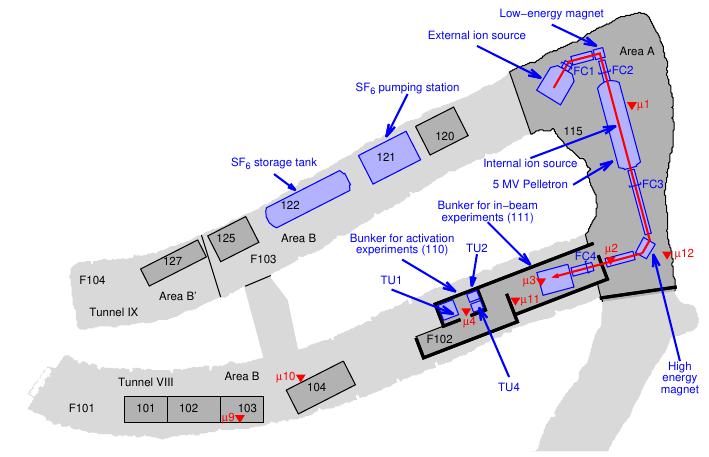}
\caption{\label{fig:Daniel_LaboratoryMap} Layout of the underground installations in Felsenkeller tunnels VIII and IX. Room numbers (three digits, for corridors with an additional leading F) are also included. Black numbers indicate technical service rooms. 
The path of the ion beam in tandem mode is given by a thick red line, and Faraday cups FC1-FC4 are marked with blue symbols. The red inverted triangles denote the approximate locations where muon flux data have been measured: prior to construction ($\upmu$1-$\upmu$4, \cite{Ludwig19-APP}) and after construction completed ($\upmu$9-$\upmu$12, this work).}
\end{figure*}

Inside tunnel VIII, there are two bunkers (rooms 110 and 111, Figure \ref{fig:Daniel_LaboratoryMap}). Bunker 110 is used for low-background offline measurements (section \ref{sec:RadioactivitySetups}), bunker 111 for in-beam measurements (section \ref{sec:InbeamSetup}).

Both bunkers are completely surrounded by 40\,cm of lower-activity concrete. The several components (sand, cement, gravel of different grain sizes, ash) of this concrete were separately studied by $\gamma$-ray spectroscopy before mixing the concrete, ensuring that the specific activity of the concrete overall was reduced with respect to the surrounding rock. For the concrete used,  specific activities of 15-17 Bq/kg $^{238}$U (1.2-1.4 ppm) and 14-18 Bq/kg $^{232}$Th (3-4 ppm) are obtained based on these analyses, about a factor of ten lower than the surrounding rock and comparable to Modane lab \cite{PALUSOVA2020106185} and to Gran Sasso hall A \cite{Wulandari04-APP}, but higher than Gran Sasso hall B.

The 40\,cm concrete thickness was chosen to ensure that the rate of $(\alpha,n)$ neutrons emitted is dominated by the low-activity concrete, not by the higher activity hornblende monzonite. A previous simulation had shown that the first 40\,cm of rock thickness dominate the $(\alpha,n)$ rate \cite{Wulandari04-APP}.

Technical rooms in tunnels VIII and IX include service rooms for the uninterruptible power supply (101), a 20\,kV transformer (102), the data link (103), the inside air unit of the air conditioning system (104), corridors (F101, F102, F103, F104), spaces for the chillers (120), for the compressed air system (125), and for the air conditioning outside air unit (127). Area A is an enclosed radiation controlled area. Areas B and B' are open tunnel but closed to public access. 

\subsection{Technical installations}

The laboratory is connected by 1 GBit/s fiberoptic cable to the main site of HZDR, and from there to the internet. Also the underground and overground sites at Felsenkeller are linked by fiberoptic cable. All of tunnel VIII and most of tunnel IX are covered by wireless LAN (eduroam). 

Both the accelerator (see below, section \ref{sec:Accelerator}) and all of the data acquisition systems are fully remote controlled from control and counting rooms at the Earth's surface, a few meters away from the tunnel entrances.

\section{Radiation background in Felsenkeller}
\label{sec:Background}

In this section, the radiation background in the Felsenkeller lab is characterized, so that dedicated shields can be developed for particular experiments. The muon (section \ref{subsec:Muon}), neutron (section \ref{subsec:Neutron}), and $\gamma$-ray (sections \ref{subsec:GammabackgroundWithoutBeam} and \ref{subsec:GammabackgroundWithBeam}) background are described, in turn.

\subsection{Muon flux and angular distribution}
\label{subsec:Muon}

The root cause of much of the background in a shallow-underground lab such as Felsenkeller is the flux of cosmic-ray muons, which are generated in the upper atmosphere. 

The effective rock overburden for the Felsenkeller laboratory has been measured prior to the construction of the present laboratory, and it was found to be 140\,meters water equivalent (m.w.e.), based on the vertical muon flux \cite{Ludwig19-APP}. At this rock overburden, an average muon energy of about 20\,GeV is expected \cite{Mei06-PRD}.

The muon flux measurements for the present laboratory have been carried out using a REGARD closed cathode chamber type muon spectrometer \cite{Barnafoeldi12-NIMA,Varga13-NIMA}. The REGARD device is characterised by high versatility and portability and is operated with argon gas with 18\,\% CO$_2$ admixture. 

Prior to the construction of the laboratory, the muon flux had been measured at four different places in tunnels VIII and IX \cite{Ludwig19-APP}. These places are labeled $\upmu$1 - $\upmu$4 in the layout map (Figure~\ref{fig:Daniel_LaboratoryMap}). The flux data have been published previously \cite{Ludwig19-APP} and are shown again here, see Table~\ref{tab:MuonIntensity}.

	\begin{table}[t]
		\begin{tabular}{|l|c|c|d{5}|d{4}|}
			\hline 
			Pos. & $\theta_{\rm max}$  & $\phi_{\rm max}$ & \multicolumn{1}{c|}{$I_{\rm max}$} & \multicolumn{1}{c|}{$J$} \\ 
			{[}Ref.] & [$^\circ$] & [$^\circ$] & \multicolumn{1}{c|}{[m$^{-2}$s$^{-1}$sr$^{-1}$]} & \multicolumn{1}{c|}{[m$^{-2}$s$^{-1}$]} \\ \hline \hline
			$\upmu$1 \cite{Ludwig19-APP} & 65(5) & 280(10) & 2.26(5) & 5.0(4)  \\
			$\upmu$2 \cite{Ludwig19-APP} & 55(5) & 280(10) & 1.80(4) & 4.6(5) \\
			$\upmu$3 \cite{Ludwig19-APP} & 55(5) & 300(10) & 1.96(5) & 4.9(4) \\
			$\upmu$4 \cite{Ludwig19-APP} & 55(5) & 280(10) & 2.66(6) & 5.4(4) \\ \hline 
			$\upmu$9 & 55(5) & 280(10) & 10.4(1) & 13.1(5) \\ 
			$\upmu$10 & 50(5) & 280(10) & 7.26(14) & 10.0(5) \\ 
			$\upmu$11 & 60(5) & 270(10) & 2.29(10) & 4.8(5) \\ 
			$\upmu$12 & 50(5) & 290(10) & 1.74(5) & 4.2(5) \\ 
			\hline \hline
		\end{tabular}
		\caption{%
			\label{tab:MuonIntensity}%
			Muon intensity. The largest differential flux $I_{\rm max}$ and the flux $J$, integrated over all angles, are listed. Measurements $\upmu$1 - $\upmu$4 have been taken before the construction of the lab inside the tunnels \cite{Ludwig19-APP}, measurements $\upmu$9 - $\upmu$12 afterwards. Only statistical errors are shown. See Figure \ref{fig:Daniel_LaboratoryMap} for the positions.} 
\end{table}

After the opening of the laboratory, using again the REGARD device, the muon flux has been studied in four additional places called $\upmu$9 - $\upmu$12 (Figure~\ref{fig:Daniel_LaboratoryMap}). These data are presented here for the first time, Table~\ref{tab:MuonIntensity}. 

In the area of the concrete bunkers, the measured total flux is $J(\upmu11)$ = 4.8$\pm$0.5 m$^{-2}$s$^{-1}$. This value is slightly lower than $J(\upmu3)$ = 4.9$\pm$0.4 and $J(\upmu4)$ = 5.4$\pm$0.4 m$^{-2}$s$^{-1}$ which had been obtained previous to the construction of the laboratory but still consistent within errors. When using the Barbouti {\it et al.} parametrization \cite{Barbouti83-JPG}, at this depth an added 40\,cm of concrete would correspond to a reduction of just 0.08 m$^{-2}$s$^{-1}$ in the total flux, consistent with the present results. The lowest total muon flux is found deep in the tunnel, $J(\upmu12)$ = 4.2$\pm$0.5 m$^{-2}$s$^{-1}$, where the rock overburden is at its maximum.

\begin{figure}[tb]
\includegraphics[width=\columnwidth]{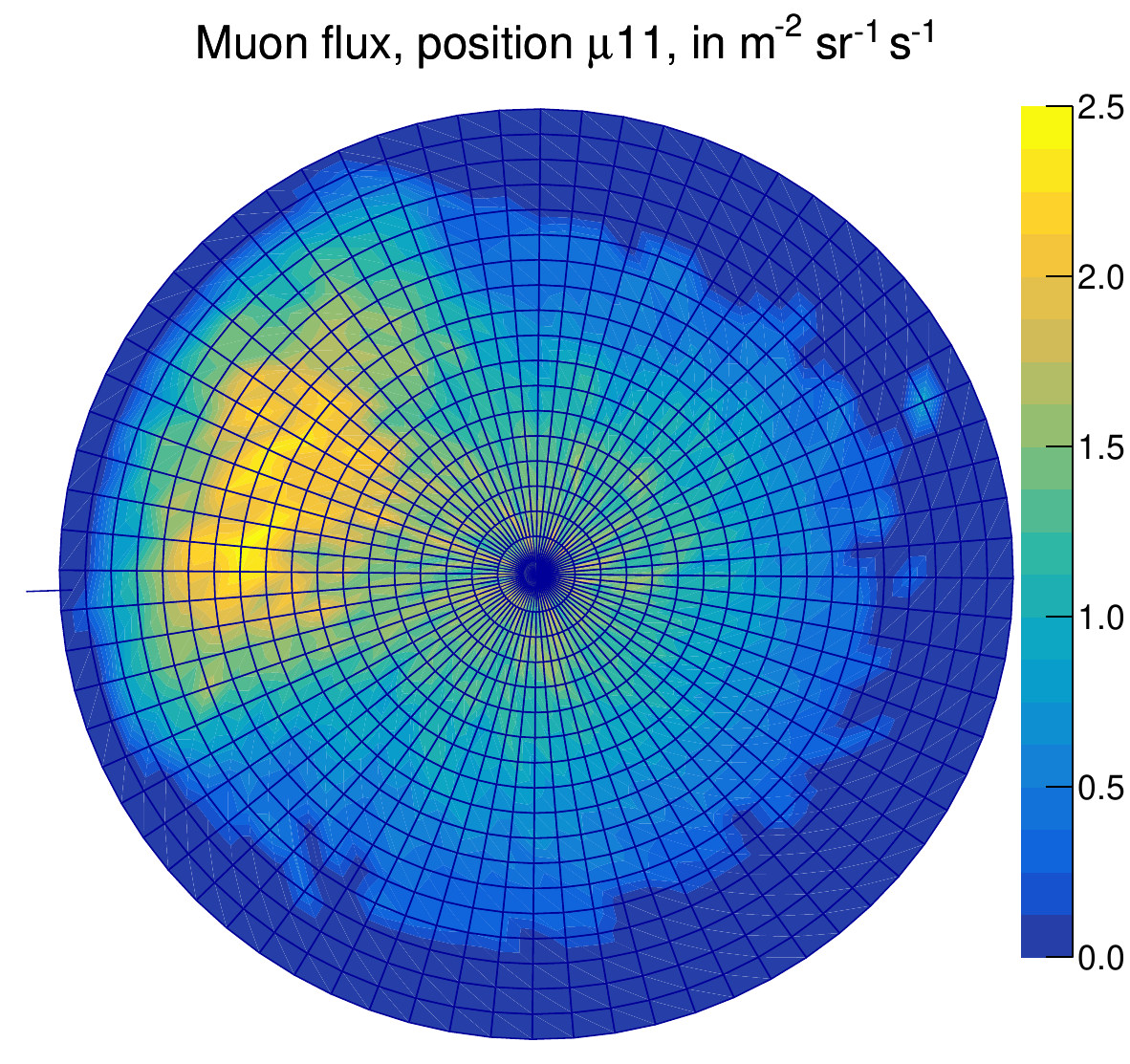}
\caption{\label{fig:Katharina_MuonAngularDistribution} Measured muon angular distribution at position $\upmu$11, in the irradiation bunker 111. North is up.}
\end{figure}

All of the measured muon angular distributions display a characteristic feature that had already been reported previously \cite{Ludwig19-APP}: The flux shows a distinct maximum in west-northwesterly direction (Figure~\ref{fig:Katharina_MuonAngularDistribution}). In this direction, the rock overburden is slightly reduced (Figure \ref{fig:BirdsEye}). The angle-integrated flux data $J$ shown in Table \ref{tab:MuonIntensity} include this maximum and thus account for this effect. 

In the concrete bunkers, at position $\upmu$11, the zenith-angle muon flux is 1.62$\pm$0.08 m$^{-2}$ s$^{-1}$ sr$^{-1}$ (Figure \ref{fig:Katharina_MuonAngularDistribution}), corresponding to a vertical rock overburden of 147 meters water equivalent. However, due to the contribution by the 60$^\circ$ maximum, for the present case it is more instructive to compare angle-integrated fluxes that take into account all effects. Assuming an angle-integrated muon flux of 190 m$^{-2}$ s$^{-1}$ at the surface of the Earth \cite{Grieder01-Book}, it is found that the angle-integrated muon flux in the measurement bunkers 110 and 111  (position $\upmu$11) is suppressed by a factor of 40$\pm$4 with respect to the surface of the Earth. 

Finally, at positions $\upmu$9 and $\upmu$10, close to the entrance of Tunnel VIII, the measured muon flux is higher, due to the features of the terrain. At these two positions, the effective rock overburdens are only 60-70 m.w.e. and these places are only used for technical installations and storage, not for experiments.

\subsection{Neutron flux and energy spectrum}
\label{subsec:Neutron}

\begin{figure}[tb]
\includegraphics[width=\columnwidth]{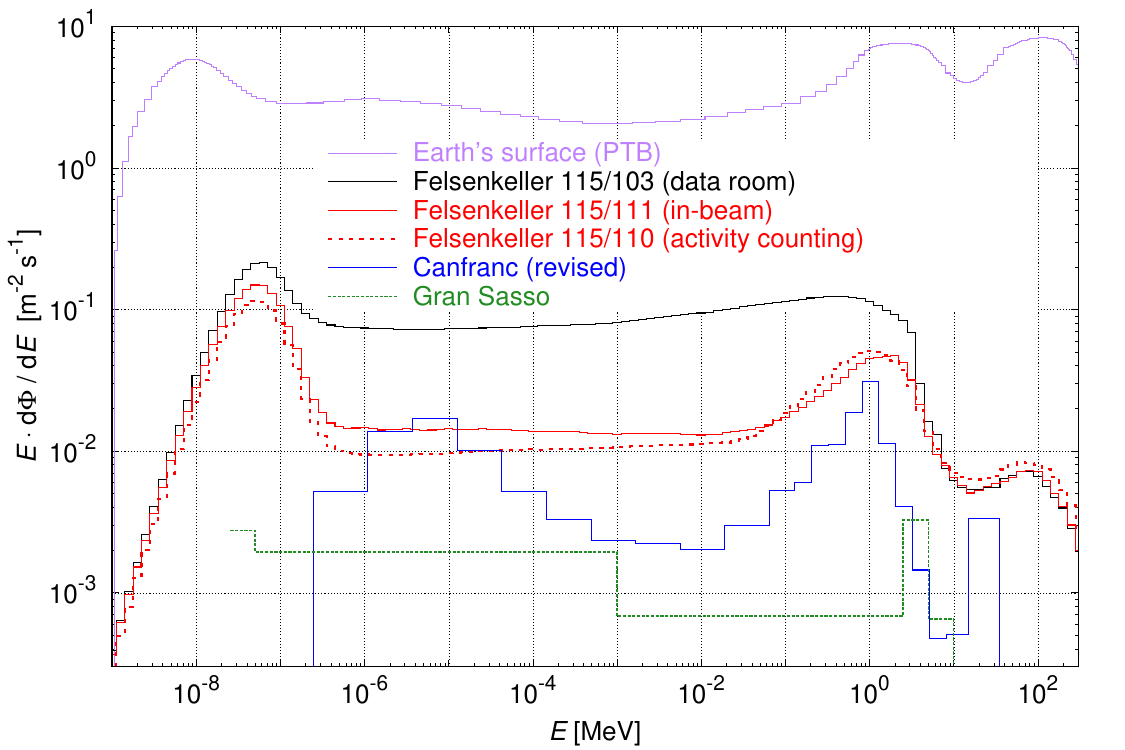}
\caption{\label{fig:nFlux} Energy spectrum of the neutron background flux at three different locations in Felsenkeller tunnel VIII. For reference, data from the Earth's surface \cite{Wiegel02-NIMA_Surface}, and from the deep-underground labs Canfranc \cite[revised as given in \cite{Jordan13-APP_Corr}]{Jordan13-APP} and Gran Sasso \cite{Belli89-NCA} are also shown. }
\end{figure}

\begin{table}[tb]
\centering
\begin{tabular}{|l|d{6}|d{7}|}
\hline
Location                    & \multicolumn{1}{c|}{Total}                 & \multicolumn{1}{c|}{$E_n>10$ MeV}              \\ 
						    & \multicolumn{1}{c|}{[m$^{-2}$\,s$^{-1}$]}  & \multicolumn{1}{c|}{[m$^{-2}$\,s$^{-1}$]}  \\ \hline \hline
Earth's surface  \cite{Wiegel02-NIMA}             & \multicolumn{1}{D{.}{}{6}|}{121.\,(6)}	        &        \\ \hline
Canfranc \cite{Jordan13-APP,Jordan13-APP_Corr} & 0.138\,(14)    &     \\ \hline
Felsenkeller VIII, 103			    & 2.19\,(15)	    &   0.033\,(7)      \\
Felsenkeller VIII, 110              & 0.61\,(3)	        &   0.029\,(3)      \\ 
Felsenkeller VIII, 111			    & 0.72\,(3)	        &   0.024\,(2)      \\
\hline
 \end{tabular}
\caption{Measured total neutron flux for Felsenkeller. For reference, the surface-based \cite{Wiegel02-NIMA} and the corrected deep underground data from Canfranc \cite{Jordan13-APP,Jordan13-APP_Corr} are also shown. }
\label{tab:TotnFlux}
\end{table}

At the depth of the Felsenkeller underground laboratory, cosmic-ray induced neutrons that make up the largest share of the neutron flux at the Earth's surface are essentially removed. Two processes contribute to the remaining underground neutron flux. The first are $(\upmu,n)$ reactions in the laboratory and its surroundings. In a study of the neutron flux in the neighboring tunnel IV, it was found previously that especially high atomic charge and high mass density materials such as steel and lead shielding may lead to problematic high fluxes of $(\upmu,n)$ neutrons \cite{Grieger20-PRD}.

The second effect contributing to the remaining neutron flux at Felsenkeller depth are $(\alpha,n)$ reactions in the surrounding rock. The natural radioactivity of the rock includes a number of $\alpha$ decays in the $^{232}$Th and $^{238}$U decay chains. These $\alpha$ particles then give rise to $(\alpha,n)$ reactions {\it in situ} in the rock. 

In a previous neutron background study performed in the nearby, different tunnel IV \cite{Grieger20-PRD}, it had been found that this effect is best mitigated using radiopure rock. The serpentinite walls of tunnel IV's MK1 chamber have specific $^{238}$U and $^{232}$Th activities of just 1.3 and 0.34\,Bq/kg, respectively. This is much lower than the specific activities of the surrounding rock, which are 130 and 170\,Bq/kg for the same decay chains, respectively \cite[and section \ref{subsec:PassiveShielding}]{Grieger20-PRD}.

The two main findings in Ref. \cite{Grieger20-PRD}, namely that at the present shallow depth steel and lead shielding may give rise to unwanted $(\upmu,n)$ neutrons and that low-radioactivity walls effectively reduce unwanted  $(\alpha,n)$, had shaped the design of the present laboratory. In particular, low-radio\-activity concrete walls of 40\,cm thickness had been adopted for the two measurement bunkers 110 and 111 (section \ref{subsec:PassiveShielding}). This shielding material combines a low $(\upmu,n)$ yield with a low $(\alpha,n)$ rate, and it also serves as enclosure of the laboratory rooms for air conditioning purposes.

After the completion of laboratory construction, including the concrete walls, the neutron flux and energy spectrum were measured {\it in situ}, using the High Efficiency Neutron Spectrometry Array\footnote{HENSA web site \url{https://www.hensaproject.org/}} setup, version HENSA \linebreak (v2018). 
This setup resembled the one used previously in Felsenkeller tunnel IV \cite{Grieger20-PRD}, with upgrades in the number of detectors and spectral resolution at low ($E_n<$ 1 eV) and high ($E_n>$ 50 MeV) energies. It consisted of eight $^3$He-filled proportional neutron counters with 10 bar gas pressure, 60 cm active length, and 1'' external diameter.
One of these $^3$He counters, called A0 \cite{Grieger20-PRD}, was used without moderator to address the thermal neutron flux. Six more were used inside polyethylene moderators of 70\,cm length and areas ranging from 4.5$\times$4.5\,cm$^2$ to 27$\times$27\,cm$^2$, called setups A1-A6 \cite{Grieger20-PRD}. The remaining detector, called setup C10, was used to address high neutron energies: C10 was moderated but also included a lead converter to enhance the detection efficiency at energies $>$50\,MeV through ($n,xn'$) multiplication reactions. 

Based on the neutron counting rates with this eight-detector array (A0-A6, C10), the energy spectrum and energy-integrated flux were obtained using spectral unfolding with FLUKA-simulated response functions, again following the previously published procedure \cite{Grieger20-PRD}.  

The energy spectra for the three locations studied here exhibit flux maxima near 1 MeV and at thermal energies (Figure \ref{fig:nFlux}). For the data room 103 near the entrance to the tunnel, an additional significant continuum at 10$^{-7}$...10$^0$ MeV was found, whereas for the two measurement bunkers 110 and 111, the flux was dominated by the thermal and 1 MeV peaks. For all three locations studied, a non-negligible flux in the region of 100 MeV was found, which may be explained by $(\upmu,n)$ neutrons. This high-energy neutron flux showed a slight decrease when moving deeper into the tunnel (Table \ref{tab:TotnFlux}), as expected given the decreasing muon flux when moving in the same direction (Table \ref{tab:MuonIntensity}).

Overall, the neutron flux integrated over all energies is attenuated by a factor of 198$\pm$14 (bunker 110) and 168$\pm$11 (bunker 111) when comparing to the surface of the earth  (Table \ref{tab:TotnFlux}). Comparing to deep-underground laboratories, one of the most extensively documented spectrally resolved measurement available is the one from Canfranc/Spain, at 2400 m.w.e. depth  \cite[revised as given in \cite{Jordan13-APP_Corr}]{Jordan13-APP}. Recently, this measurement was updated through a long-term experiment using HENSA \cite{Orrigo22-EPJC}, confirming previous results \cite{Jordan13-APP,Jordan13-APP_Corr}.
The present total flux in Felsenkeller is a factor of 4-5 higher than the reported total flux at Canfranc, possibly due to the predominant contribution  of $(\alpha,n)$ neutrons produced in the concrete walls there. 
It is noted that the setup used in Ref. \cite[revised as given in \cite{Jordan13-APP_Corr}]{Jordan13-APP} does not include a thermal neutron detector, so the integral flux may still increase when taking this into account. A new measurement campaign at Canfranc using the HENSA detector array is currently ongoing \cite{Orrigo21-JPCS}, taking thermal neutrons into account and showing the expected slight increase with respect to Ref. \cite[revised as given in \cite{Jordan13-APP_Corr}]{Jordan13-APP}. 
The comparison to Gran Sasso is more difficult due to the limited number of energy bins used in that work \cite{Belli89-NCA}. It is noted that at Gran Sasso, a new wide energy range neutron flux measurement campaign utilizing HENSA has commenced.


\subsection{$\gamma$-ray background without incident ion beam}
\label{subsec:GammabackgroundWithoutBeam}

The background observed in $\gamma$-ray detectors at Felsenkeller has been studied in details previously \cite{Szucs12-EPJA,Szucs15-EPJA,Szucs19-EPJA,Turkat23-APP}. In the following it is discussed for high-purity germanium (HPGe) detectors with a cosmic-ray muon veto. The cosmic-ray muon veto in those cases was given by the escape suppression veto detector, which is a BGO detector surrounding the HPGe detector on the sides, in a typical configuration for accelerator-based measurements where the front face of the HPGe is left without veto in order to allow $\gamma$ rays from the reaction target to reach the HPGe unobstructed \cite{Szucs12-EPJA,Szucs15-EPJA,Szucs19-EPJA}.

Two different cases are relevant: First, $E_\gamma <$ 3\,MeV, which is relevant both for radionuclide analyses and in-beam $\gamma$-ray measurements and where abundant data are available. Second, $E_\gamma >$ 3\,MeV, which is only relevant for in-beam $\gamma$-ray measurements and where data is scarce.

\begin{figure}[tb]
\includegraphics[width=\columnwidth,trim=0mm 0 12mm 0,clip]{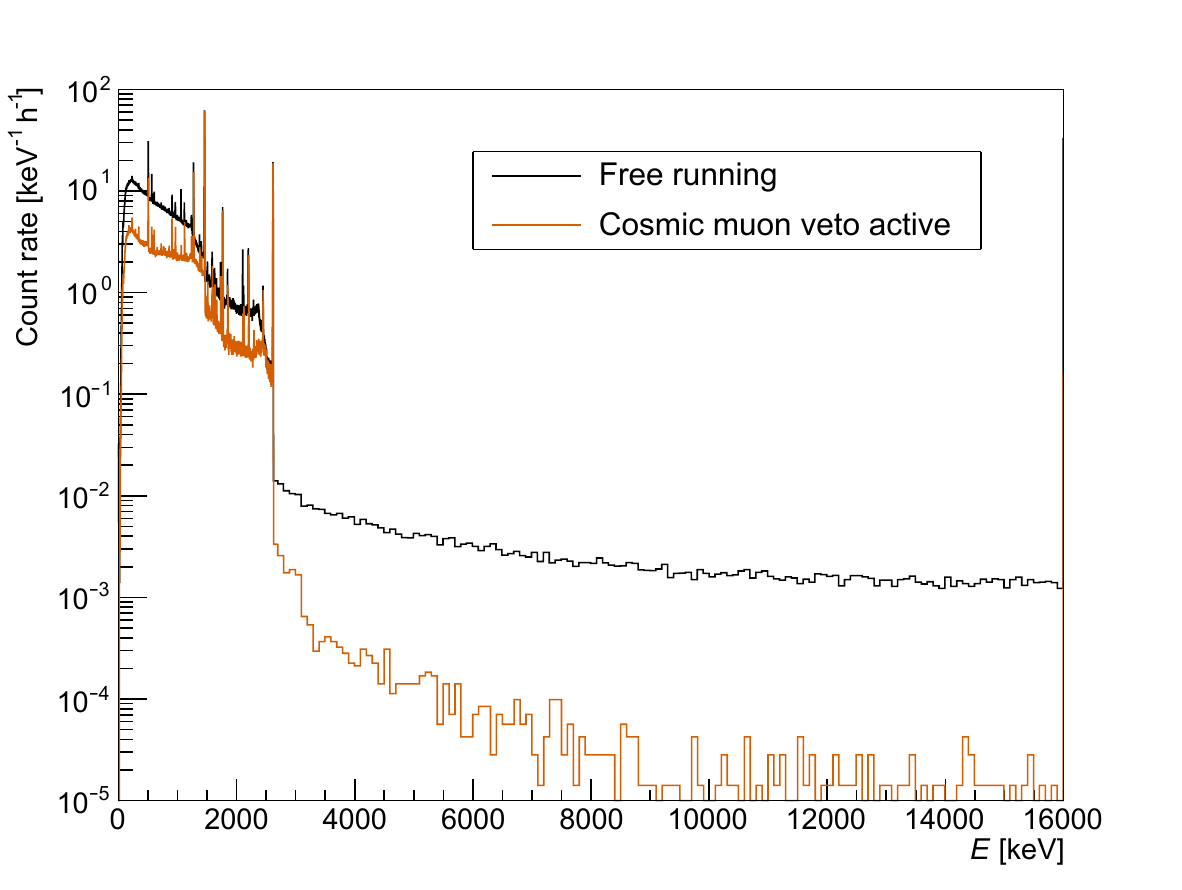}
\caption{\label{fig:Ortec90BG} Typical no-beam $\gamma$-ray background spectra. The histograms are based on a 90\,$\%$ HPGe detector positioned inside a BGO detector and a lead castle enclosure.}
\end{figure}

\begin{figure}[tb]
\includegraphics[width=\columnwidth]{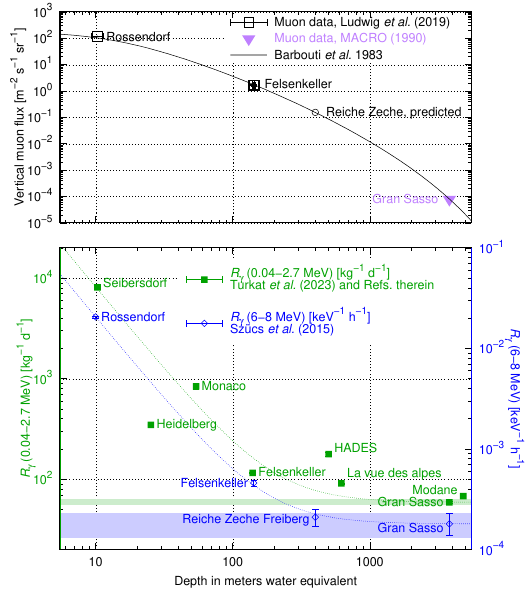}
\caption{\label{fig:MuonVertDiffInt} Top panel: Measured \cite{Ludwig19-APP,Ahlen90-PLB} and calculated \cite{Barbouti83-JPG} muon fluxes as a function of rock overburden in m.w.e. --- Bottom panel: Two different background count rates. Left axis and green symbols, rates $R_\gamma$\,(0.04-2.7\,MeV) in units of kg$^{-1}$d$^{-1}$ \cite[and references therein]{Turkat23-APP}. Right axis and blue symbols, rates $R_\gamma$\,(6-8\,MeV) in units of keV$^{-1}$h$^{-1}$ \cite[and references therein]{Szucs15-EPJA}. The dashed lines are just to guide the eye. }
\end{figure}

For $E_\gamma <$ 3\,MeV, highly sophisticated shielded setups are constructed which include, for laboratories below 1000 m.w.e. depth, an active cosmic-ray veto. A quantity that is frequently adopted for background comparisons is the counting rate in the 0.04-2.7 MeV $\gamma$-energy region, normalized to the germanium mass of the detector used \cite{Laubenstein04-Apradiso}. The TU1 radioactivity counting setup of Felsenkeller shows a background of 1982$\pm$3 kg$^{-1}$d$^{-1}$, and with muon veto it is reduced to 113$\pm$1 kg$^{-1}$d$^{-1}$ \cite{Turkat23-APP}. This background is just a factor of two higher than the detector with the lowest reported background in the literature, the deep-underground GeMPI detector set at Gran Sasso \cite{ACKERMANN2023110652}.

The $E_\gamma >$ 3\,MeV region is important for in-beam $\gamma$ ray measurements in nuclear astrophysics, due to the high $Q$ value of a number of the most important reactions. Prominent examples are $^{14}$N($p,\gamma$)$^{15}$O (hydrogen burning, \cite{Marta08-PRC,Wagner18-PRC}) with a $Q$ value of 7.3 MeV and $^{2}$H($p,\gamma$)$^{3}$He (Big Bang nucleosynthesis, \cite{Mossa20-Nature,Turkat21-PRC}) with a $Q$ value of 5.5 MeV. For comparisons in these energy regions, the same HPGe detector has been used successively at different overground and underground laboratories \cite{Szucs10-EPJA,Szucs12-EPJA,Szucs15-EPJA}, with a cosmic-ray veto. The energy region of 6-8 MeV has been chosen, which is unaffected by detector-intrinsic $\alpha$-activities and $\gamma$-ray pileup effects. Here, the Felsenkeller-based background is (4.6$\pm$0.3)$\times10^{-4}$ keV$^{-1}$h$^{-1}$ (cosmic muon veto active), a factor of 2.6$\pm$0.7 times higher than at Gran Sasso \cite{Szucs15-EPJA}. For illustration, the effect of the cosmic muon veto at Felsenkeller is shown in Figure \ref{fig:Ortec90BG} for a different detector. 

The background counting rates for Felsenkeller and selected other overground and underground laboratories are displayed in Figure \ref{fig:MuonVertDiffInt}. In the Figure, in the top panel it is seen that the muon flux varies over six orders of magnitude between the surface of the Earth and deep-underground labs such as Gran Sasso \cite{Ludwig19-APP,Ahlen90-PLB}. The bottom panel shows that the background rates in characteristic regions of interest for $\gamma$-ray detectors. 

When considering the $\gamma$-ray background indices (Figure \ref{fig:MuonVertDiffInt}, bottom panel), the lowest values for each background region are found at the deep-underground Gran Sasso lab \cite{Laubenstein04-Apradiso,Szucs10-EPJA}. These values are marked as shaded regions in Figure \ref{fig:MuonVertDiffInt}. The Felsenkeller background values are 2.0-2.6 times higher, as discussed above.


\subsection{$\gamma$-ray background with incident ion beam}
\label{subsec:GammabackgroundWithBeam}

\begin{figure}[tb]
\includegraphics[width=\columnwidth,trim=0mm 0 0mm 0,clip]{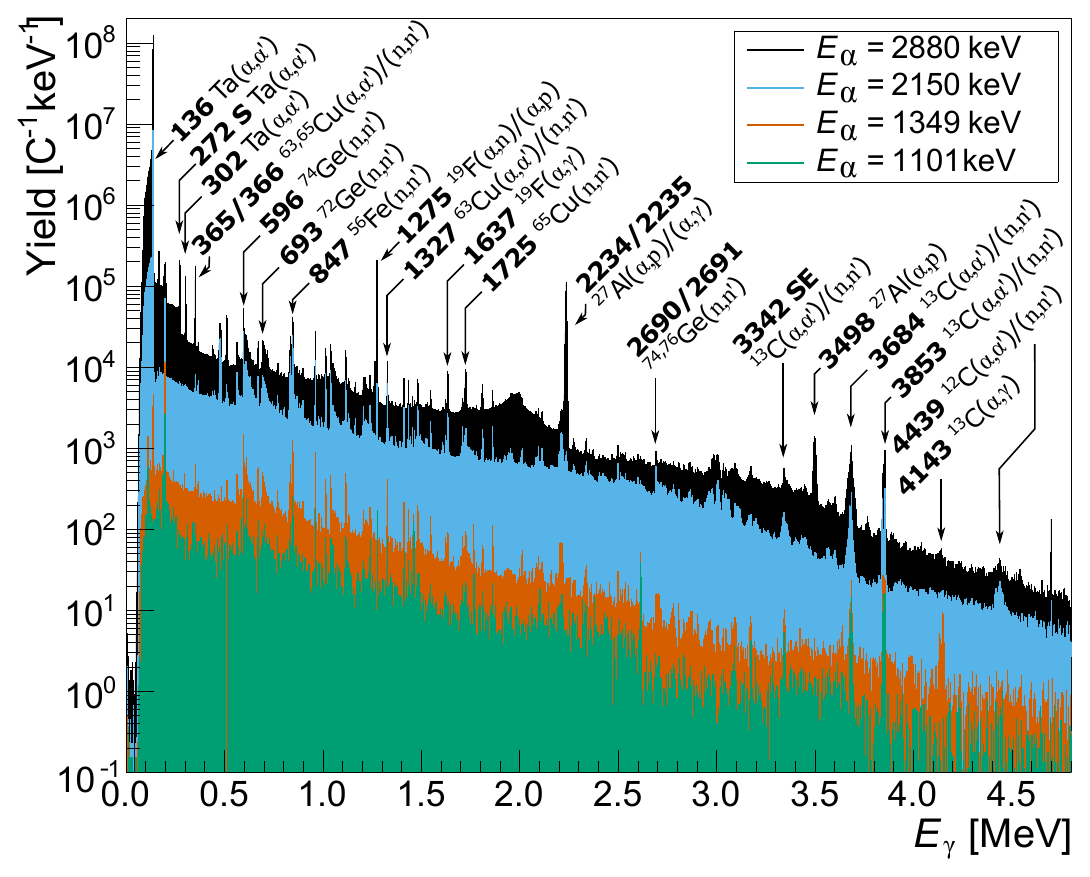}
\caption{\label{fig:InbeamBG} Typical in-beam $\gamma$-ray background spectra with incident $\alpha$ beams of $E_\alpha$ = 1101-2880 keV on a blank tantalum plate. The histograms are based on the sum of three 60\,$\%$ HPGe crystals positioned in close geometry at 90$^\circ$ with respect to the beam axis. For all spectra, the no-beam background has been subtracted.}
\end{figure}

The $\gamma$-ray background induced by an incident ion beam, defined as undesired $\gamma$-rays that are emitted either promptly or as a result of activation caused by the ion beam, depends strongly on the beam species, the experimental setup, and the beam energy. Here, as an illustration only the case of a $^{4}$He$^+$ beam incident on chemically cleaned, blank tantalum plate is discussed, as a function of beam energy (Figure \ref{fig:InbeamBG}). The Ta plate is frequently used as a target backing. The vacuum in the target area was always kept in the 10$^{-8}$ \dots 10$^{-7}$ hPa range in order to limit impurities.

At low energies, $E_\alpha$ = 1101-1349 keV (green and orange spectra in Figure \ref{fig:InbeamBG}), the $\alpha$-beam induced background is quite limited and mainly given by $(\alpha,\gamma)$ and  $(\alpha,\alpha')$ reactions on $^{13}$C included in carbon impurities deposited on top of the blank Ta plates. Furthermore, some limited neutron-induced features appear, namely characteristic triangular shapes near 596 and 693 keV. They are due to neutrons created by the $^{13}$C($\alpha,n$) reaction.

At high energies, $E_\alpha$ = 2150-2880 keV (blue and black spectra in Figure \ref{fig:InbeamBG}), due to the higher Coulomb barrier penetrability the $^{13}$C($\alpha,n$) rate and, hence, the rate of neutron-induced effects increases considerably. Furthermore, some transfer and capture reactions on nuclei of higher charge numbers start to play a role, namely \linebreak $^{19}$F($\alpha,p$)$^{22}$Ne, $^{27}$Al($\alpha,p$)$^{30}$Si, and $^{27}$Al($\alpha,\gamma$)$^{31}$P. The latter two reactions are significant because they indicate unwanted beam losses on aluminum. This may be beam scattered in a collimator hitting the aluminum-made piece of beam line hosting it during this test run. 

For $E_\alpha > $ 2350 keV (black spectrum in Figure \ref{fig:InbeamBG}), i.e. above the threshold of the $^{19}$F($\alpha,n$)$^{22}$Na reaction, there is another factor of 2-3 increase in the neutron yield, due to the trace amounts of fluorine present in the tantalum backing. 

The quantitative importance of the ion beam induced background can best be gauged by taking again an integral over a large region of interest that is typically free of effects caused by natural radionuclides. When using the integral in the 3.1-4.8 MeV $\gamma$-ray energy region, the counting rate for the spectra (after subtraction of the no-beam background) per incident charge varies over orders of magnitude: 
From 
4.7$\times$10$^2$ and
2.4$\times$10$^3$ C$^{-1}$ ($E_\alpha$ = 1101 and 1349 keV, respectively) at the lower two beam energies, to
6.4$\times$10$^4$ and
2.6$\times$10$^5$ C$^{-1}$ ($E_\alpha$ = 2150 and 2880 keV, respectively) at the higher beam energies where there are many $\alpha$-induced neutrons.

\section{The ion accelerator system at Felsenkeller}
\label{sec:Accelerator}

In this section, the accelerator (section \ref{subsec:Pelletron}) and its external (section \ref{subsec:SputterIonSource}) and internal (section \ref{subsec:TerminalIonSource}) ion sources are described. The beam line components are listed, as well (section \ref{subsec:Beamline}).

\subsection{5\,MV Pelletron accelerator}
\label{subsec:Pelletron}

The accelerator is a model 15SDH-2 Pelletron ion accelerator with 5\,MV nominal terminal voltage, manufactured by National Electrostatics Corporation, USA. The accelerator had originally been used as a $^{14}$C accelerator mass spectrometry system in York, UK (Xceleron Inc.) \cite{Young08-York} and was moved to Dresden, Germany, in 2012. 

The high voltage terminal is located in the center of a pressure tank, which is filled with up to 6.5\,bar of SF$_6$ gas for insulation purposes during operation. In order to maintain the good insulation properties, the gas is constantly cooled and dried. The high-voltage generation is based on the Van-de-Graaf principle realized by the mechanical charging system with two chains, which are made of individual metal pellets. With positive upcharging on the grounded side and negative downcharging on the high voltage terminal a charging capacity of 250\,$\upmu$A for 50\,Hz operation of the motor can be achieved.

The high voltage stability is ensured by a Terminal Potential Stabilizer (TPS) system. Two capacitive pickoff (CPO) plates measure the high-frequency changes of the terminal voltage. The general stability can be observed by either using a generating voltmeter inside the pressure tank or a slit system after the bending magnet behind the accelerator if ion beam is extracted. An automated stabilization based on the evaluated signals is then realized by a corona probe inside the tank. After maintenance work usually a voltage conditioning is needed to guarantee stable operation, especially for higher voltages. The voltage gradient of both sides of the high voltage terminal can be influenced by the use of a shorting rod.

When the accelerator is used together with the external ion source, the negative ion beam from the sputter source (see section \ref{subsec:SputterIonSource} below) propagates through the low-energy side of the Pelletron. It then passes a gas stripper unit that is operated in recirculating mode with a two-stage pumping scheme. The main stripper cell is a 57\,cm long tube filled with nitrogen, typical pressure 3$\times$10$^{-3}$ - 3$\times$10$^{-2}$ hPa, corresponding to 4$\times$10$^{15}$ - 4$\times$10$^{16}$ cm$^{-2}$ gas thickness.
In this mode a total gain in potential energy by $(1+Q)e\cdot U$ is achieved, depending on the terminal voltage $U$ and the charge state $Q$ selected after the stripper. 

Downstream of the gas stripper unit and still inside the high voltage terminal, there is a terminal ion source that can be used when the external ion source is switched off, see section \ref{subsec:TerminalIonSource} below. 

The accelerator control, as well as the control of the ion sources and other beam line components are realised by a custom-built LabVIEW\footnote{National Instruments, https://www.ni.com} based control system that is controlled via fiberoptic cable from the surface-based accelerator control room.

\subsection{Sputter ion source}
\label{subsec:SputterIonSource}

\begin{figure}
    \centering
    \includegraphics[width=\columnwidth]{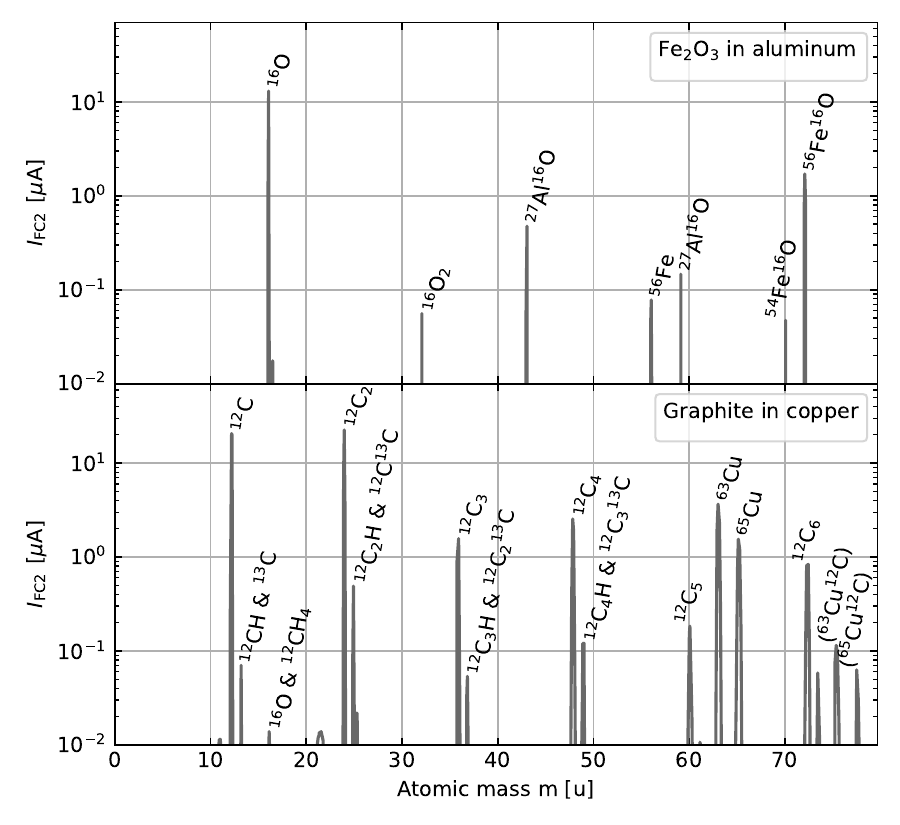}
    \caption{Magnetic scan of the beam from the sputter ion source for iron oxide (Fe$_2$O$_3$, top panel) and graphite (bottom panel) cathode materials in two different holders. The current on Faraday cup FC2 downstream of the low-energy magnet is plotted as a function of the ion mass, assuming single electric charge.}
    \label{fig:Magnetscan_SputterSource}
\end{figure}

When the accelerator is used in its original tandem mode, the negatively charged beam is supplied by a sputter ion source with 134 samples (cathodes). The ion source is of type 134 MC-SNICS (multi cathode source of negative ions by cesium sputtering) by NEC. Its construction, operation, and performance at its previous place of installation in York/UK has been described elsewhere \cite{Loger99-AIPCP}. At Felsenkeller, so far carbon and oxygen beams have been developed, but in principle many other beam species are possible, as well. 

The 134 MC-SNICS uses cesium sputtering to produce negatively charged ions from the sample cathode. The cesium is housed in a stainless steel cylinder, heated by a coil in the so-called cesium oven to 140-160$^\circ$C, and reaches the ioniser through a heated transport line. The generated positively charged cesium ions are then accelerated by a potential difference of about 5\,kV onto the cathode. The ions sputtered out of the cathode are then extracted by 20-40 kV bias. 

In a typical case with carbon cathode, about $200\,\upmu$A of negative current are achieved on Faraday cup FC1 (see Figure \ref{fig:Daniel_LaboratoryMap} for the position of the Faraday cups). This beam is then electrically deflected by 45$^\circ$ and transported to a 1.2\,T electromagnet, the so-called low-energy (LE) magnet. 

The magnetic scans of the beam on FC2, located just behind the LE magnet shows the usual pattern of a sputter ion source. 
For oxygen beams, here two different cathode materials have been tested: Fe$_2$O$_3$ and WO$_3$. For Fe$_2$O$_3$ cathodes, in addition to a 12 $\upmu$A of mass-16 beam ($^{16}$O$^-$), there are also oxygen molecules ($^{16}$O$_2^-$) and beams including iron and aluminum  (Figure \ref{fig:Magnetscan_SputterSource}, upper panel). The aluminum oxide (mass 43) is probably due to imperfect focusing of the cesium beam leading to some sputtered out aluminum structural material, probably the cathode holder. Typically, for Fe$_2$O$_3$ (WO$_3$) cathodes about 90\% (70\%) of the electrically measured beam on FC1 is found as $^{16}$O beam on FC2, respectively, meaning that the desired $^{16}$O beam dominates over unwanted parasitic beams.

Carbon beams were generated from graphite cathodes. They show a "molecular" beam $^{12}$C$_2^-$ that is almost as strongly represented as the main beam, mass-12 ($^{12}$C$^-$). In addition, also molecules containing up to at least six carbon atoms are found, as well as possible traces of the $^{13}$C isotope (1.1\% natural abundance). The graphite scan also shows some hydrocarbon molecules and $^{63}$Cu and $^{65}$Cu ions from the copper cathode holder (Figure \ref{fig:Magnetscan_SputterSource}, lower panel).

\subsection{Terminal ion source}
\label{subsec:TerminalIonSource}

\begin{figure}[tb]
\includegraphics[width=\columnwidth]{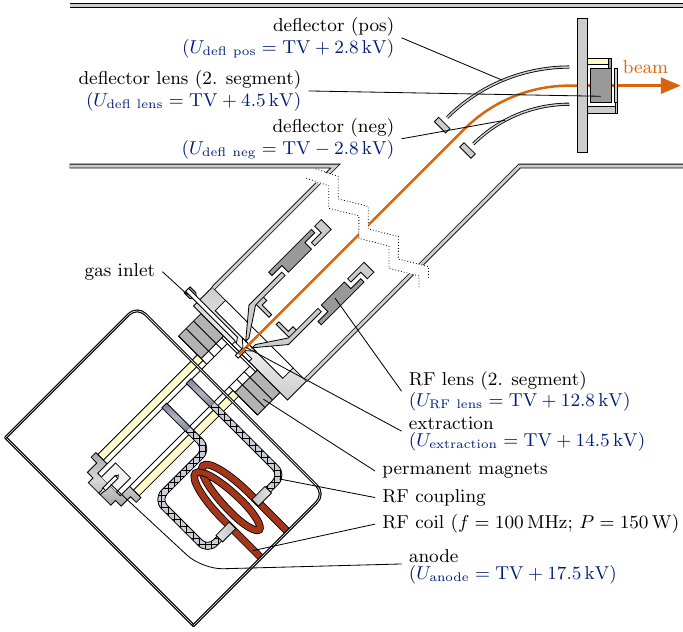}
\caption{\label{fig:Deflector} Schematic drawing of the terminal-mounted, internal ion source including the electrostatic deflector. Indicated are voltages for a long-term alpha beam extraction of $\sim30\,\upmu$A. See text for details.}
\end{figure}

Immediately downstream of the gas stripper unit, there is a positive radio frequency (RF) ion source, manufactured by NEC, mounted directly on the high voltage terminal. 
The gas supply to the source is adjusted by a remote controlled needle valve. An RF oscillator with a fixed output power of 150\,W generates 100\,MHz high frequency, which is coupled into the source glass to generate a plasma from the used gas. This effect is intensified by a permanent magnet around the exit of the source glass. The source is mounted under an angle of 30$^\circ$ relative to the beam tube (Figure~\ref{fig:Deflector}).
Via an anode at the end of the RF ion source glass and a cylindrical extraction channel, the ions can be extracted from the plasma. The extraction is followed by an einzel lens, also made by NEC, the deflector and another custom-built einzel lens that is directly connected to the deflector. A typical set of voltages for the operation of the ion source is included in Figure~\ref{fig:Deflector}.

 A custom-built electrostatic deflector is used to bend the beam from its original 30$^\circ$ angle to 0$^\circ$ for the acceleration. The deflector includes two curved, parallel plates that can be individually biased and a casing. The individual parts are connected with L-shaped ceramic parts in order to inhibit parasitic currents between different parts of the deflector. The upper plate has a hole, through which the beam of the external sputter ion source can pass. Thereby a quick change between the internal and the external source is possible, without the need of additional maintenance.
 
In  this configuration, the ion source has been well tested for both proton and helium irradiations with typical target currents of 10-20\,$\upmu$A. With minor adjustments, successful long-term runs were done with stable beam extraction over weeks continuously.

\begin{figure}[tb]
\includegraphics[width=\columnwidth]{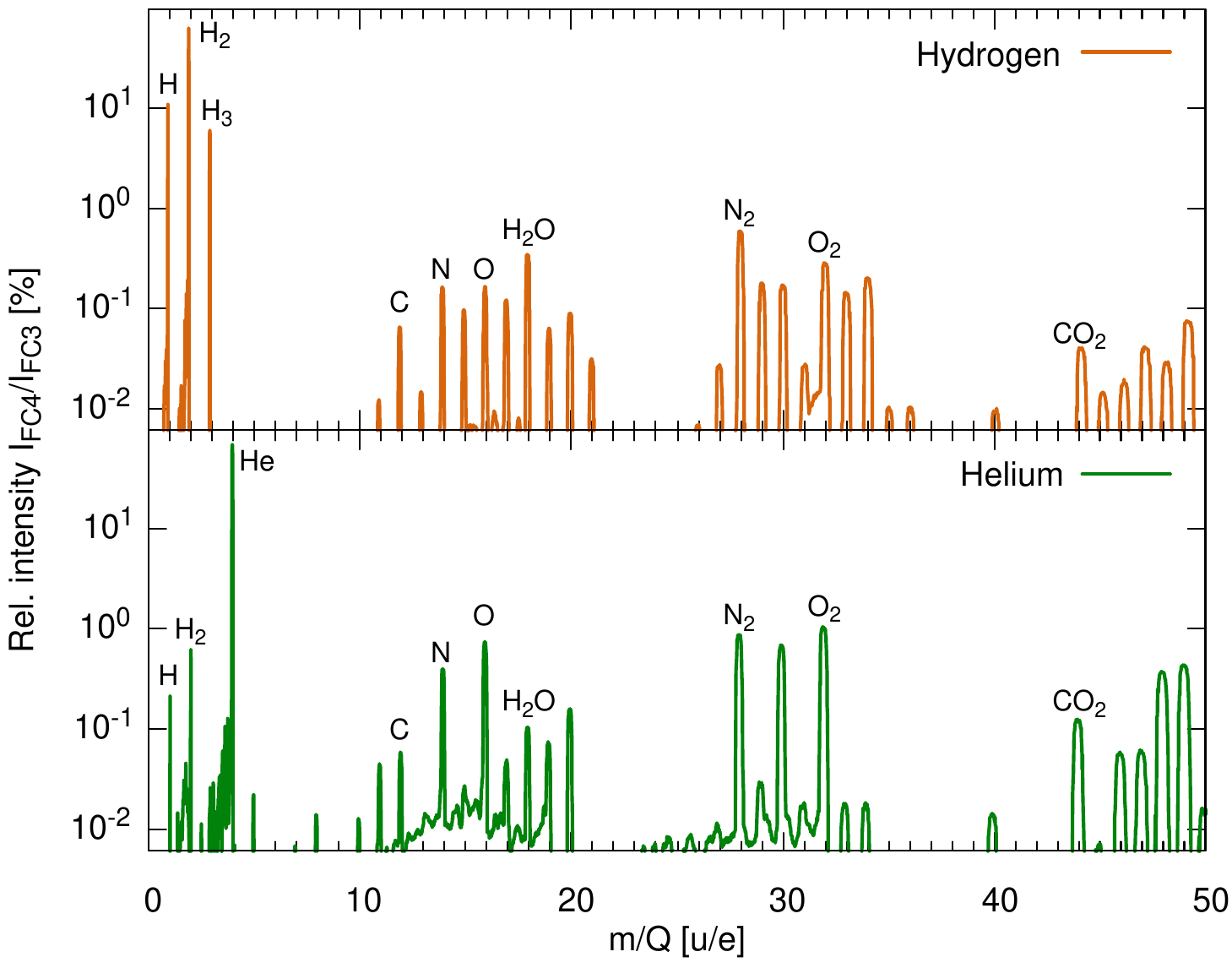}
\caption{\label{fig:magnetscan_combined} Magnetic analysis of the ion beam provided by the internal RF ion source, as measured on FC4 after the HE magnet. The top (bottom) panel shows the data for hydrogen (helium) gas filled into the RF source.}
\end{figure}

To analyze the composition of the beam, magnet scans have been performed, measuring the current on the Faraday cup FC4 behind the high-energy magnet. 
The analysis of two typical magnet scans done with the internal ion source can be found in Figure~\ref{fig:magnetscan_combined}, where the resulting beam composition with hydrogen (top) and helium (bottom) as source gas are shown.

For both hydrogen and helium gas, the spectrum shows a dominant peak for the gas of interest: for hydrogen gas, it is molecular hydrogen H$_2^+$. For helium gas, it is singly-charged atomic helium He$^+$. Additionally, after a source maintenance the emission spectra show a number of background ions from carbon, carbohydrates, air (nitrogen and oxygen), and water. All these unwanted additional items decrease over time during operation. For the helium beam, there is a memory due to previous operation of the source with hydrogen, as evidenced by the H$_2^+$ and H$^+$ peaks (Figure \ref{fig:magnetscan_combined}).

\subsection{Beam line}
\label{subsec:Beamline}

In the following, the various components of the beam line are discussed in the order of the propagation of the ion beam.

The vacuum system at Felsenkeller is divided into six sections along the beam line: (1) sputter source, (2) low-energy magnet and beam line, (3) high-energy beam line, (4) high-energy magnet, (5) target room, and (6) target. The sections are separated by five electropneumatically operated vacuum shutters labeled shutter~1~to~5 according to the preceding vacuum section. In each section, ultra-high vacuum at typical pressures of $10^{-9} \dots 10^{-7}$\,hPa monitored by a Pfeiffer PKR 251 compact full range gauge is maintained by a Pfeiffer HiPace~700 turbomolecular pump, backed by a Pfeiffer ACP40 multi-stage roots pump. In case of a failure of the ACP40 fore-pump, it is separated from the turbo pump by a Pfeiffer solenoid actuated angle valve AVC 025 MA.  

The intensity of the ion beam is measured by four Faraday cups. FC1 behind the sputter source, FC2 behind the low-energy magnet, FC3 behind the Pelletron accelerator, and FC4 behind the high-energy magnet. The precise locations of FC1-FC4 are given in Figure \ref{fig:Daniel_LaboratoryMap}. Each cup consists of a $-200$\,V biased molybdenum aperture, and a conical tantalum collector. Each cup is mounted on a remote controlled electropneumatic actuator moving it in and out the beam line.

The position and distribution of ions within the beam are measured by six NEC Model BPM-8 series beam profile monitors. 
The output signals from the beam profile monitors are read out by digital oscilloscopes, calibrated in distance units, and displayed on the operator screen. The BPM generated beam shape files can be downloaded in digital form for further analysis.

After passing, subsequently, the sputter ion source (section \ref{subsec:SputterIonSource}), FC1, and Shutter 1, there is a 45$^\circ$ electrostatic deflector (NEC electrostatic analyzer, 300 mm radius). 

After the deflector, the ion beam passes a 90$^\circ$ double focusing electromagnet with 457\,mm right-hand bending radius and up to 1.2\,T field. The induction is controlled by changing the current between 0 and 120\,A at a nominal voltage of 34.5 V. The dipole acts as first mass-over-charge filter. Subsequently, the ions pass BPM2, FC2 and are injected into the low-energy side of the 5\,MV Pelletron accelerator.

After leaving the Pelletron, the positively charged ions pass a 90$^\circ$ electromagnetic dipole magnet with a bending radius of 1270\,mm and up to 1.5\,T field. Acting as a second mass-over-charge filter, it separates different charge states in case of the use of the external sputter ion source (section \ref{subsec:SputterIonSource}) and the terminal electron stripper (section \ref{subsec:Pelletron}) or different ion species in case of the use of the internal radio-frequency source (section \ref{subsec:TerminalIonSource}).

Downstream of the high-energy magnet, there are slits that may be used for the terminal voltage stabilization, alternative to the generating voltmeter. Further, there are beam profile monitor BPM5, Faraday cup FC4, BPM6, a magnetic quadrupole doublet lens made by NEC (option 3, MQD3), and the connection to the target chamber, which is described below in section \ref{sec:InbeamSetup}.

\subsection{Energy calibration of the Pelletron}
\label{subsec:EnergyCalibration}

\begin{figure}
    \centering
    \includegraphics[width=\columnwidth]{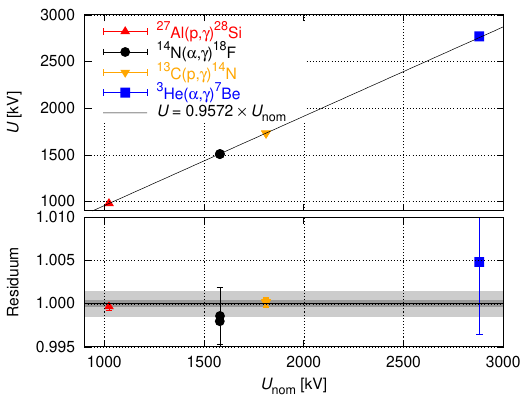}
    \caption{Energy calibration of the accelerator based on nuclear reactions. In the residual plot, the dark (light) shaded regions indicate the uncertainty when not including (including) uncertainties introduced by accelerator maintenance between calibration and experiment, respectively. See text for details.}
    \label{fig:EnergyCalib}
\end{figure}

The terminal voltage of the accelerator is measured by a generating voltmeter (GVM) made by NEC. The reading of the GVM signal is logged every few seconds in the automatic accelerator log and here called the nominal terminal voltage $U_{\rm nom}$. In order to obtain the true physical voltage $U$, a calibration factor $C = U/U_{\rm nom}$ has to be determined. The beam energy $E_{\rm beam}$ is then given by
\begin{eqnarray}
E_{\rm beam} & = & q_{\rm in} ( U_{\rm SNICS} + U) + q_{\rm out} (U + U_{\rm RF}) = \\ \nonumber
        & = & q_{\rm in} (U_{\rm SNICS} + C \, U_{\rm nom}) + q_{\rm out} \left( C \, U_{\rm nom} + U_{\rm RF} \right)
\end{eqnarray}
where $q_{\rm in}$ is the charge number of the injected ion beam (0 for single-ended mode and 1 for tandem mode), $q_{\rm out}$ is the charge number of the ion beam after the accelerator, $U_{\rm SNICS}$ is the accelerating potential of the SNICS sputter ion source and $U_{\rm RF}$ the accelerating potential by the internal radio-frequency ion source. If the accelerator is operated in tandem mode, $U_{\rm RF}$ is zero.. 

The calibration factor $C$ was then obtained using several different methods, with consistent results. For simplicity, the calibration was only carried out in single-ended mode, i.e. $q_{\rm in}$ = 0. 

First, the current on the low and high energy columns of the terminal was measured and multiplied with the column resistances of 109 and 114 G$\Upomega$, respectively. This method was only precise to 5\% due to the difficulty of measuring such high resistances, but it gives an important consistency check for the other results. 

Furthermore, the resonance scan technique using well-known resonances in radiative capture reactions was applied (Figure \ref{fig:EnergyCalib}). For hydrogen beam, the $E_p$ = (991.76 $\pm$ 0.02) keV resonance in $^{27}$Al($p,\gamma$)$^{28}$Si \cite{NDS28-2013} and the $E_p$ = (1747.6 $\pm$ 0.9) keV resonance in $^{13}$C($p,\gamma$)$^{14}$N \cite{Ajzenberg13_15_91-NPA} were used. For helium beam, the doublet of states at excitation energies $E_x$ = 5603.38 and 5604.86 keV in $^{18}$F was used \cite{Tilley_18-19}, corresponding to a resonance energy of $E_\alpha$ = (1528.8$\pm$1.0) keV with the error band covering both levels. 

In addition, direct-capture $\gamma$ rays from the $^3$He($\alpha,\gamma$)$^7$Be reaction ($Q$ value 1587.14$\pm$0.07 keV) have been used at a nominal beam energy of $E_\alpha$ =  2.880 MeV. Here, the location of the $\gamma$ ray in the spectrum has been determined and used to trace back the physical terminal voltage.

The resonance scans were repeated for each new experimental campaign. Deviations by up to 0.15\% were found after an in-tank maintenance, possibly due to slight shifts in the generating voltmeter calibration after work in the pressure tank. 

The final calibration factor was then determined to be
\begin{equation}
    C = 0.9572 \pm 0.0015
\end{equation}
with the uncertainty dominated by the possible shifts due to accelerator maintenance (light grey shaded area in Figure \ref{fig:EnergyCalib}). If instead an ad hoc calibration is performed, the typical error is about three time smaller, $\pm$0.0005  (dark grey shaded area in Figure \ref{fig:EnergyCalib}). 

\section{Irradiation station for in-beam $\gamma$-spectrometry}
\label{sec:InbeamSetup}

The ion beam generated by the acceleration system (section \ref{sec:Accelerator}) is then transported to room 111, the bunker for in-beam experiments (Figure \ref{fig:Daniel_LaboratoryMap}). The height of the beam line is 1.6\,m above the floor. The room has total dimensions of 8.9\,m (length) $\times$ 3.2\,m (width) $\times$ 3.0\,m (height). Within these parameters, various kinds of target and detection station can be mounted. At the time of this writing, a solid target irradiation station is mounted on the beam line and described here as an example (section \ref{subsec:SolidTarget}). 

\subsection{Solid target irradiation station}
\label{subsec:SolidTarget}

In case of solid target experiments there are four different types of target holders available at the Felsenkeller laboratory, here called W55, W0, N01, N02. 
All four are compatible with targets that are deposited atop a metallic foil (typically 0.22\,mm thick tantalum) of 27\,mm diameter. This foil is thick enough to completely stop the ion beam in all possible scenarios at Felsenkeller, and thin enough to efficiently transport the heat deposited by the stopping beam to its back side. 
There, cooling water of typically 15\,$^\circ$C (holders W55 and W0) or the end of a liquid-nitrogen cooled copper finger (holders N01 and N02) with typically -100\,$^\circ$C temperature with beam (-196\,$^\circ$C without beam) are present to transport the heat away.

The first of the two water-cooled holders, W55, is designed for $55^{\circ}$ angle between beam direction and target normal and has been used in a number of previous surface-based experiments \cite{Schmidt13-PRC,Schmidt14-PRC,Depalo15-PRC,Wagner18-PRC}. W90 is a modified copy of W55, where the target normal is parallel to the direction of the incident ion beam. 

The LN$_2$-cooled holder N01 is connected via a copper finger to a dewar and designed for experimental regimes of high,  $>$20 W, beam power on the target. The optimized holder N02 includes less passive materials attenuating emitted $\gamma$-rays from the nuclear reaction under study. Both N01 and N02 keep the target perpendicular to the incident ion beam.

\begin{figure}[tb]
\includegraphics[width=\columnwidth]{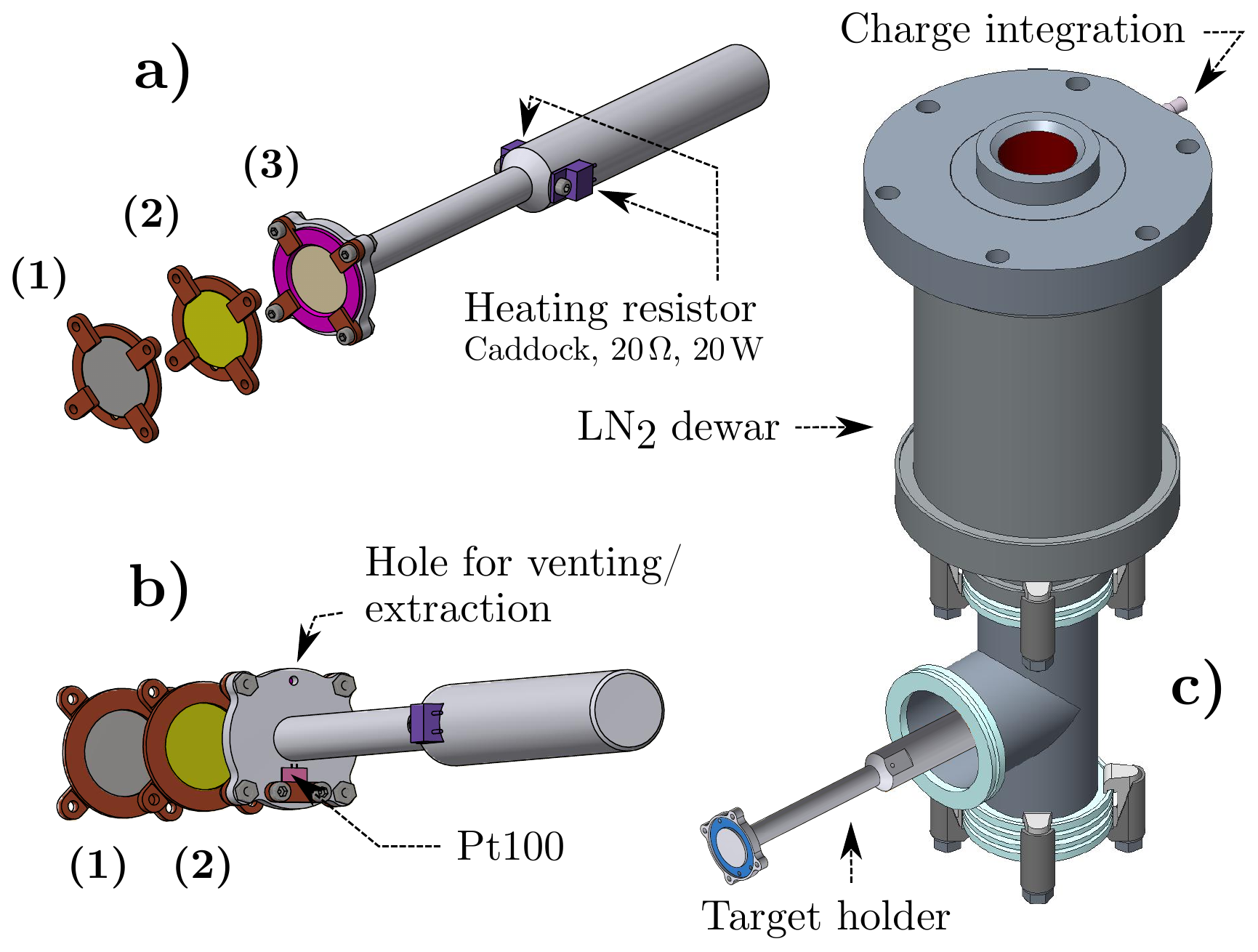}
\caption{\label{fig:Targetholder} Target holder N02 used for liquid nitrogen cooling. The ion beam is incident from the left side in the picture.  a)~Front view. b)~Back view. c)~Overview. See text for details.}
\end{figure}

As an example, target holder N02 is shown in Figure \ref{fig:Targetholder}. In addition to the standard target for irradiation (option (1) in Figure  \ref{fig:Targetholder}, panels a and b), also two different types of radionuclide calibration standards can be mounted on the target holder (options (2) and (3) in the Figure). 

For target holders N01 and N02, the effective target temperature is monitored using a Pt100 thermoresistor mounted on the back side of the target holder (Figure  \ref{fig:Targetholder}, panel b). In order to efficiently warm up the target holder prior to venting the beam line for a target change, two 20\,$\Omega$ heating resistors are mounted at the sides of the cold finger (Figure  \ref{fig:Targetholder}, panel a). The target holder and its cold finger are electrically insulated from the outside of the setup, in order to allow for the electrical measurement of the beam current via a charge integrator (Ortec 439). Further details can be found in thesis work \cite{Turkat23-PhD}.

\subsection{$\gamma$-ray detection efficiency calibration}

For the calibration of the $\gamma$-ray detection efficiency, a number of thin radionuclide standards can be mounted in the setups described in section \ref{subsec:SolidTarget}. These standards, provided by the German metrology institute PTB (Physikalisch-Technische Bundesanstalt Braunschweig), are usually calibrated to 1.0-1.5\% precision (95\% confidence interval) in activity. 

Standards available at the laboratory include $^{22}$Na ($E_\gamma$ = 511, 1275\,keV),  $^{57}$Co ($E_\gamma$ = 14, 122, 136\,keV), $^{60}$Co ($E_\gamma$ = 1173, 1333\,keV), $^{65}$Zn ($E_\gamma$ = 1115\,keV), $^{88}$Y ($E_\gamma$ = 898, 1836\,keV), $^{133}$Ba ($E_\gamma$ = 53 - 384\,keV), $^{137}$Cs ($E_\gamma$ = 662\,keV), and $^{241}$Am. 

The $\gamma$-ray detection efficiency curve can then be extended up to high energy, 10.6 MeV, using the well-known resonances at 0.992 and 0.278\,MeV in the $^{27}$Al($p,\gamma$)$^{28}$Si and $^{14}$N($p,\gamma$)$^{15}$O reactions, respectively \cite{Anttila77-NIM,Zijderhand90-NIMA,Daigle16-PRC}.

\subsection{Available detectors and calibration samples}

A number of high-purity germanium (HPGe) detectors are available and can be arranged around the interaction chamber. All of the detectors can be fitted with bismuth germanate (BGO) escape suppression shields, available at the laboratory. All of them have a coaxial central contact, also in case of the hexagonal crystals.

A list of the in-beam HPGe detectors that are used as of this writing, is given in Table \ref{tab:HPGe}. The EB17, EB18, MB1, and MB2 cluster consist of altogether 20 hexagonal Euroball \cite{Wilhelm96-NIMA} detectors. For the MB1 and MB2 triple clusters, three Euroball detectors each are arranged in a Miniball triple cryostat. The EB17 and EB18 septuple clusters contain seven Euroball detectors each \cite{Eberth96-NIMA}. They are kindly supplied by the GAMMAPOOL. Further details are found in thesis work \cite{Turkat23-PhD}.

The no-beam background for the MB1 and MB2 detectors has already been published \cite{Szucs19-EPJA} and is also described in section \ref{sec:Background}.

\begin{center}
\begin{table}
\caption{
\label{tab:HPGe} 
In-beam HPGe detectors available at Felsenkeller as of this writing. See text for details.}  
\begin{tabular}[ht]{| l | l | l | l |} 
 \hline
 Name & Detector & Type &{Rel. eff.}  \\ 
   &  &  &  \multicolumn{1}{c|}{} \\ 
  \hline\hline
Can60   &   Single  & Cylindrical, p-type & 1$\times$60\% \\ 
Ortec90     &   Single  & Cylindrical, p-type &    1$\times$88\% \\
MB1     &   3-cluster   & Hexagonal, n-type & 3$\times$60\%  \\
MB2     &   3-cluster  & Hexagonal, n-type &   3$\times$60\%  \\
EB17    &   7-cluster   & Hexagonal, n-type &   7$\times$60\%  \\ 
EB18    &   7-cluster   & Hexagonal, n-type &  7$\times$60\%  \\ 
 \hline
\end{tabular}
\end{table}
\end{center}

\section{Radioactivity measurement setups}
\label{sec:RadioactivitySetups}

In addition to the accelerator-based measurements (bunker 111 in Figure \ref{fig:Daniel_LaboratoryMap}), there are several radioactivity measurement setups in bunker 110. For the sake of discussion, these detectors and setups are called TU1-TU7 and listed in Table \ref{tab:TUBunker}. They include five HPGe detectors and two silicon drift detectors (SDD) for X-ray measurements. In addition, a number of other detectors not listed in the Table are available: large (1$\times$1 m) plastic scintillator paddles as muon veto, as well as LaBr$_3$, CeBr$_3$, GaGG, NaI, CsI, CZT, and PIPS detectors.

\begin{center}
\begin{table}
\caption{
\label{tab:TUBunker}
Offline counting detectors in bunker 110. For the HPGe detectors, the relative efficiency $\eta_{\rm rel}$ is listed. For the SDD detectors (sensitive for 1-20\,keV X-rays), the active area $A$ is given.}  
\begin{tabular}[ht]{| l | l | p{45mm} | r | } 
 \hline
 Name & Type & Specifics & $\eta_{\rm rel}$  \\ 
  \hline\hline
 TU1 & HPGe & Coax, p-type, see Ref. \cite{Turkat23-APP} & 163\% \\ 
 TU2 & HPGe & Borehole, well-type, 4$\pi$ for small samples & 54\%  \\
 TU3 & SDD & $A$ = 80\,mm$^2$ & \\
 TU4 & HPGe & Coax, n-type & 41\% \\
 TU5 & SDD & $A$ = 175\,mm$^2$  &  \\ 
 TU6 & HPGe & Coax, p-type & 37\% \\ 
 TU7 & HPGe & Coax, n-type, 12\,$\upmu$m mylar entrance window & 38\% \\ 
 \hline
\end{tabular}
\end{table}
\end{center}

\subsection{Ultra-low background HPGe detector TU1}
\label{subsec:TU1}

\begin{figure}[tb]
\includegraphics[width=\columnwidth]{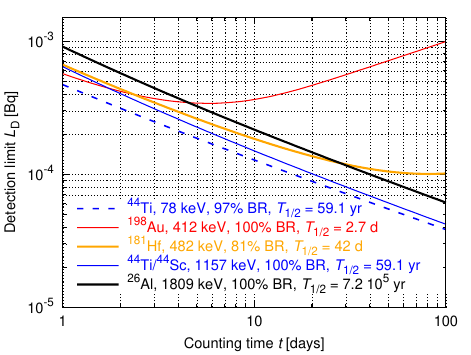}
\caption{\label{fig:TU1_DetectionLimit} Detection limit $L_{\rm D}$, at 90\% confidence level, as a function of counting time $t$ for a point-like sample directly positioned on the end cap of TU1 for five selected $\gamma$ rays. The assumed width of the region of interest ranged from 4 keV ($E_\gamma$ = 78 keV) to 10 keV  ($E_\gamma$ = 1809 keV).}
\end{figure}

The TU1 detector is a coaxial p-type HPGe detector with a relative efficiency of 163\,\% (Table \ref{tab:TUBunker}), 3.06\,kg crystal mass, and 574 cm$^3$ crystal volume. The detector and shielding have been described in a dedicated publication \cite{Turkat23-APP}. The resulting background counting rate in the 40\,keV $\leq$ $E_\gamma$ $\leq$ 2700\,keV $\gamma$-ray energy range is 113(1)\,kg$^{-1}$d$^{-1}$ \cite{Turkat23-PhD,Turkat23-APP}, just 1.9 times higher than the lowest-background detector reported in the worldwide literature \cite{Laubenstein04-Apradiso}.

The detection limit achievable with this detector depends on sample geometry including self-absorption inside the sample, $\gamma$-ray branching and energy, half life, measurement time, and radioactive contaminations in the sample other than the radionuclide to be studied. For nuclear astrophysics activation experiments, a point-like sample and negligible self-absorption can usually be assumed. 

In Figure \ref{fig:TU1_DetectionLimit}, some examples for detection limits achievable are plotted as a function of measurement time. For the relatively short-lived nuclides $^{198}$Au and $^{181}$Hf, the detection limit is essentially given by statistics and shows an optimal value of 100-300 $\upmu$Bq when the sample is counted for 2-3 half lives (red and orange curves in Figure \ref{fig:TU1_DetectionLimit}). 

For longer-lived nuclides, the detection limit keeps improving over time up to the highest counting time of 100 days studied here (blue dashed, blue solid, and black lines in Figure \ref{fig:TU1_DetectionLimit}), with values of 40-60 $\upmu$Bq found for 100 days counting time for the examples studied here.

A web-based online tool is available to compute detection limits for this detector \cite{Webpage24_TU1}.

The innermost volume of the passive shielding allows sample sizes with a diameter of 150 mm and a height of 39 mm. When removing two upper copper disks from the inner shielding, the possible height of samples can be increased to 89 mm. Sample holders made of oxygen-free radiopure (OFRP) copper are available for various geometries and distances.

\subsection{Borehole detector TU2}
\label{subsec:TU2}

The TU2 detector is a borehole HPGe detector (SAGe Well, S-ULB configuration) from Mirion/Canberra with an active volume of 218\,cm$^{3}$. The crystal well has 21\,mm diameter and is 40\,mm deep. The cryostat well diameter is 16\,mm, so that cylindrical samples of less than 16\,mm diameter and 40\,mm height may be placed in the well. 

For counting nuclear astrophysics point-like samples, the borehole geometry is not used, because those samples usually present in form of disks (typical diameter 27\,mm) that are too large to fit in 21\,mm the borehole. For these samples, appropriate holders  are available on which the samples can be placed at distances of 2.5\,mm, 30\,mm and 70\,mm, on top of the end cap, respectively. 

The detector is mounted within a passive shielding consisting of 5\,cm of OFRP copper and 15\,cm of low-activity lead. An additional anti-radon box is surrounding this setup. The innermost volume of the passive shielding allows sample sizes of 
98\,x\,98\,x\,150\,mm placed on top of the end cap and is continuously flushed with nitrogen from the LN$_2$ dewar of the detector. 

The TU2 background counting rate without a muon veto is currently $R=8495(12)$\,kg$^{-1}$d$^{-1}$ when integrating over the $\gamma$-ray energy range 40-2700\,keV. This value is four times higher than the TU1 value without a muon veto \cite{Turkat23-APP}, and 75 times higher than the TU1 value with a muon veto (section \ref{subsec:TU1}). The analysis of the offline spectrum shows some residual natural radionuclide background in the TU2 setup, presumably inside the detector and preamplifier assembly. Due to the still too high background for the TU2 detector, currently no muon veto is foreseen for this setup.

\subsection{Other detectors for radioactivity measurements}

In addition to the TU1 and TU2 detectors discussed in sections \ref{subsec:TU1} and \ref{subsec:TU2}, three smaller HPGe detectors are available for offline counting: TU4, TU6, and TU7 (Table \ref{tab:TUBunker}). They range in relative efficiency from 37-41\%. TU4 is hosted in a commercial graded shield by von Gahlen with 10\,cm thick lead shielding. The other detectors may be moved to ad hoc shielding.

For the detection of X-rays, two silicon drift detectors (SDD) are available (Table \ref{tab:TUBunker}). Their active areas are 109 and 170 mm$^2$, respectively. The shielding for these detectors is still under development.

\section{Summary and outlook}
\label{sec:Summary}

A new underground accelerator laboratory has recently opened in tunnels VIII and IX of the Felsenkeller area, Dresden, Germany. 

In this work, the physical characteristics of the laboratory and its detailed background characterization have been described. The cosmic-ray muon flux has been found to be reduced by a factor of 40 when compared to the surface, and detailed angular flux maps have been developed for altogether eight positions in the underground facility. Similarly, the ambient neutron flux and energy spectrum has been measured at three different locations in Felsenkeller. Inside the measurement bunkers 110 and 111, the energy-integrated neutron flux is 170-200 times lower than at the surface of the earth.

The background in germanium $\gamma$-ray detectors has been studied, as well, and with an active muon veto the $\gamma$-ray background is just two times higher than deep underground. With beam, the precise background depends on the beam type and energy. 

The ion accelerator system has been described and characterized in detail, including both ion sources, the beam line, and the energy calibration of the accelerator. 

The irradiation station for in-beam $\gamma$-ray spectroscopy has been described using one particular experiment as an example. The various setups for offline counting have been presented, and detection limits in the 10$^{-4}$ Bq range have been calculated.

For outside scientific users, there is the possibility to apply for free of charge usage of the ion accelerator infrastructure  by submitting a scientific project to an independent outside scientific advisory board for the laboratory. In addition, currently European Union - funded Transnational Access is available via the ChETEC-INFRA project \footnote{Web page \url{www.ChETEC-INFRA.eu}, 2021-2025.}. 

The new Felsenkeller underground ion accelerator is now completely characterized and open to outside users.

\subsection*{Acknowledgments}
Enlightening discussions with Luis Mario Fraile (Universidad Complutense de Madrid, Spain) and technical support by Bernd Rimarzig, Maik Görler, Toralf Döring, and Andreas Hartmann (HZDR) are gratefully acknowledged. 
 --- This work has made use of GAMMAPOOL resources.  
Financial support has been provided by Deutsche Forschungsgemeinschaft DFG, Germany (INST 269/631-1 \linebreak FUGG, TU Dresden Institutional Strategy "support the best",  ZU123/21-1, and BE4100/4-1), by the Konrad-Ade\-nau\-er-Stiftung, Germany, by the European Union \linebreak (ChETEC-INFRA, 101008324), by the Spanish Ministerio de Economía y Competitividad under Grants FPA2017-83946-C2-1-P and FPA2017-83946-C2-2-P, and by Deut\-scher Akademischer Austauschdienst (DAAD). E.M. acknowledges an Alexander von Humboldt postdoctoral fellowship. 


\end{document}